\documentclass[journal = jpcc, manuscript = article]{achemso}
\usepackage{setspace}
\setkeys{acs}{articletitle = true, etalmode=truncate, maxauthors = 10}
\usepackage{graphicx}
\usepackage{amsmath,amssymb,amsfonts,bm}
\usepackage{mathtools}
\usepackage[colorlinks,citecolor=blue,urlcolor=black]{hyperref}
\usepackage{hyperref}
\usepackage{braket}
\author{Naseem Ud Din}
\affiliation{Department of Chemistry, Wayne State University, Detroit, Michigan 48202, USA}
\author{Zhen-Fei Liu}
\email{zfliu@wayne.edu}
\affiliation{Department of Chemistry, Wayne State University, Detroit, Michigan 48202, USA}
\title{Anisotropy of the Optical Properties of Pentacene:Black Phosphorus Interfaces}

\begin{document}

\begin{abstract}
Black phosphorus (BP) is a layered material with anisotropic properties. We study interfaces formed by a pentacene monolayer adsorbed on monolayer BP, a prototypical system for BP surface passivation. We place the pentacene monolayer along the zigzag and armchair directions of the BP substrate, respectively, to examine the anisotropy of the heterogeneous interfaces. We perform first-principles $GW$ plus Bethe–Salpeter equation ($GW$-BSE) calculations to determine the quasiparticle and optical properties. To quantitatively analyze the anisotropy of the optical properties, we develop a general computational scheme to decompose the interface excitons into different contributions. We find a distinct charge-transfer exciton formed when the monolayer pentacene is placed along the armchair direction, and discuss how the anisotropy of each component is modulated by the interface. Our results shine a light on the understanding of the BP surface passivation via molecular adsorption and provide a benchmark for future experimental and computational studies.
\end{abstract}

Black phosphorus (BP) has attracted significant attention as a low-dimensional layered material.\cite{XWJ2014,LWHX2014,CWZR2016} Perhaps its most prominent feature is the in-plane anisotropy, manifested in the optical absorption,\cite{MTXW2016,TSLY2014,LRCJ2014,WJST2015,LRKC2016} electrical and thermal transport,\cite{FFSY2014,QKHY2014,HHWC2015,LMDD2015,LYSY2015} plasmons,\cite{LRWX2014,LA2016,CAD2017} electron-photon and electron-phonon interactions,\cite{LHHL2016,LZQD2015} among others. These properties make BP suitable for various applications, such as optical linear polarizer,\cite{MTXW2016} photodetector,\cite{YLAL2015} and gas sensor \cite{KFC2014}. Furthermore, BP features direct electronic band gaps that can be tuned by the number of layers,\cite{TSLY2014,CZZ2014} promising for applications in field-effect transistors \cite{LYYG2014,LNZL2014} and photonics.\cite{GZLW2015,XWXD2014}

One challenge in the development of BP-based functional devices is to address its degradation in ambient conditions.\cite{CVPI2014, ISZC2015, FGFP2015} This instability can be largely mitigated by passivation and encapsulation using other layered materials \cite{WWJC2014,AVTW2015,LSWZ2017,DOKY2015} or molecular adsorbates \cite{WNLW2017,YLJC2016,RWWY2016}. It is therefore of paramount interest to quantitatively understand the effect of the passivation or functionalization on the properties of BP. In fact, the doping of atoms and the adsorption of molecules on the BP surface have been the object of many prior studies.\cite{KFC2014,RWWH2016} Among others, Ref. \citenum{QJL2017} showed that the electronic and optical properties of few-layer BP could be effectively tuned by other extended substrates and layered materials via many-body effects. For the interfaces formed between BP and molecular adsorbates, additional questions arise. Monolayers of organic molecules could form along either the armchair or the zigzag direction of the BP, giving rise to potentially anisotropic behaviors in addition to the intrinsic anisotropy of pristine monolayer BP. We will try to address the following two questions in this work: How would different orientations of the molecular layer modulate the anisotropy of the BP surface? On the other hand, how would the anisotropy of the BP surface affect the properties of the molecular adsorbate? 

Here, we leverage state-of-the-art first-principles $GW$-BSE formalism ($G$: Green's function; $W$: screened Coulomb interaction; BSE: Bethe-Salpeter equation) within the framework of many-body perturbation theory \cite{H1965, HL1986,RL2000} to study the electronic and optical properties of the prototypical interfaces formed between a monolayer of pentacene molecules and the monolayer BP. For a full account of the anisotropy in such interface systems, we consider two orientations of the molecule monolayer adsorbed on the BP surface, where the long axis of the pentacene molecule is aligned along either the armchair or the zigzag direction of BP. For each interface structure, we consider the optical absorption of light polarized along either the armchair or the zigzag direction of BP, which gives rise to the optical anisotropy of pristine monolayer BP. The $GW$-BSE approach systematically improves over local and semi-local density functional calculations of quasiparticle and optical properties of heterogeneous interfaces,\cite{KMH2014,CQ2018,FL2021} thanks to its capturing of the long-range dielectric screening, crucial in an accurate description of the interfacial level alignment.\cite{NHL2006,TR2009} In addition to standard $GW$-BSE calculations, we further develop a new computational analysis tool to decompose the excited states of the interface (as calculated from BSE) into different contributions, useful in understanding the nature of each peak in the absorption spectra and unraveling the interface effect in modulating the excitonic properties of each individual component. 

\begin{figure}[h]
\centering
\includegraphics[width=4in]{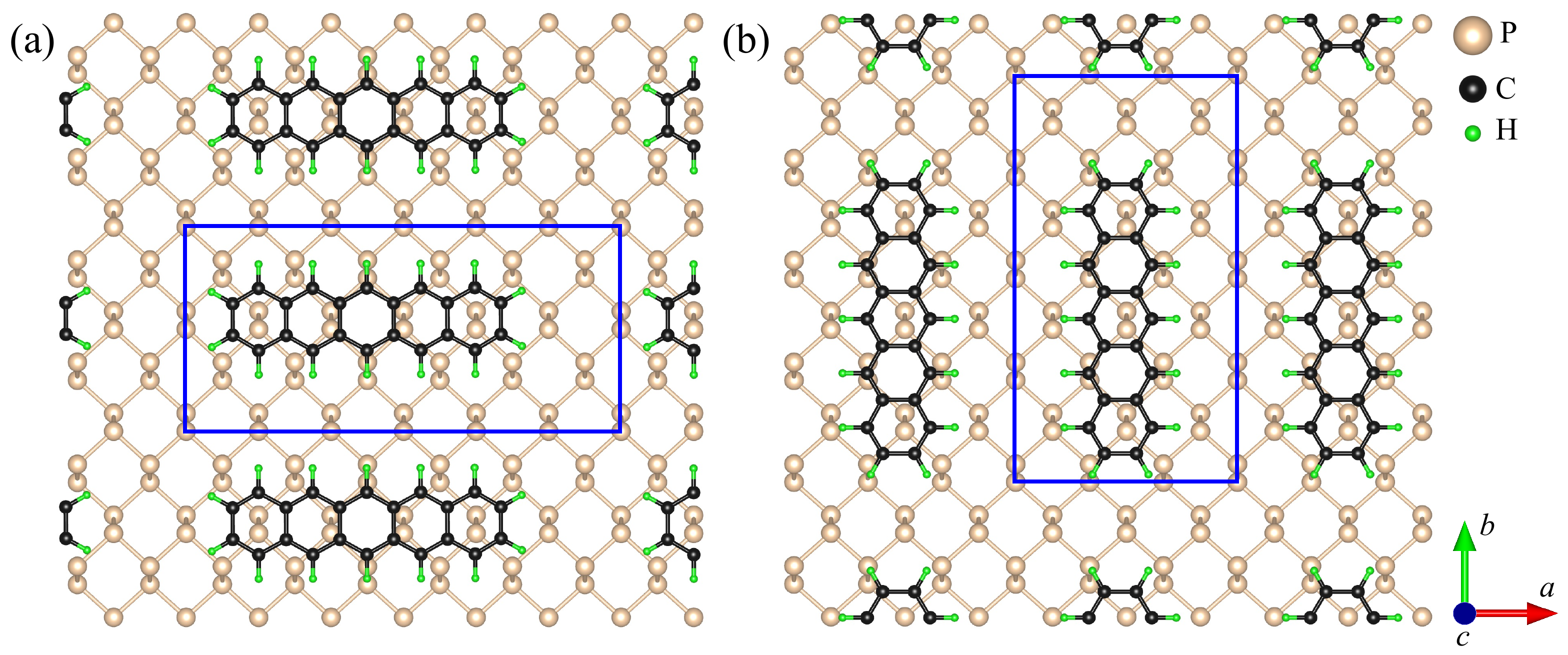}
\caption{Schematic representations of the pentacene:BP (a) zigzag interface and (b) armchair interface. The blue boxes denote the interface simulation cell.}
\label{fig:stru}
\end{figure}

We start by relaxing the lattice parameters and atomic coordinates of the monolayer BP unit cell structure using the Perdew-Burke-Ernzerhof (PBE) functional\cite{PBE1996} as implemented in the Quantum ESPRESSO package.\cite{GABB2017} The calculation uses the optimized norm-conserving Vanderbilt (ONCV) pseudopotentials\cite{SG2015}, a $\mathbf{k}$-mesh of 12$\times$8$\times$1, and a kinetic energy cutoff of 50 Ry. The resulting in-plane lattice parameters are $a=3.30$ \AA~and $b=4.63$ \AA, in good agreement with previously reported values\cite{LNZL2014} of 3.35 \AA~and 4.62 \AA, respectively. After the monolayer BP unit cell relaxation, we build two pentacene:BP interfaces. In the first (second) one, the interface consists of a 6$\times$2 (3$\times$4) supercell of monolayer BP and the long axis of the pentacene molecules is aligned along the zigzag (armchair) direction of the BP. For simplicity, we call this system the ``zigzag interface'' (``armchair interface'') in this work. In both cases, the interface cell is 30.0 \AA~along the $c$ direction. During the relaxation of the interface, the substrate BP atoms are kept fixed in their relaxed monolayer positions and the coordinates of the adsorbate molecule are fully relaxed until all residual forces are below 0.05 eV/\AA. This strategy allows us to focus on the electronic interactions across the interface by ignoring the difference in the substrate geometries. The relaxations use the vdw-DF-cx functional\cite{BH2014}, a $\mathbf{k}$-mesh of 2$\times$4$\times$1 (4$\times$2$\times$1) for the zigzag (armchair) interface, and a kinetic energy cutoff of 70 Ry. The resulting molecular monolayer is nearly flat on the BP surface, and we found an adsorption height of about 3.28 \AA~(3.38 \AA) for the zigzag (armchair) interface. Fig. \ref{fig:stru} shows the relaxed structures of the two interfaces.

\begin{figure}[h]
\centering
\includegraphics[width=4in]{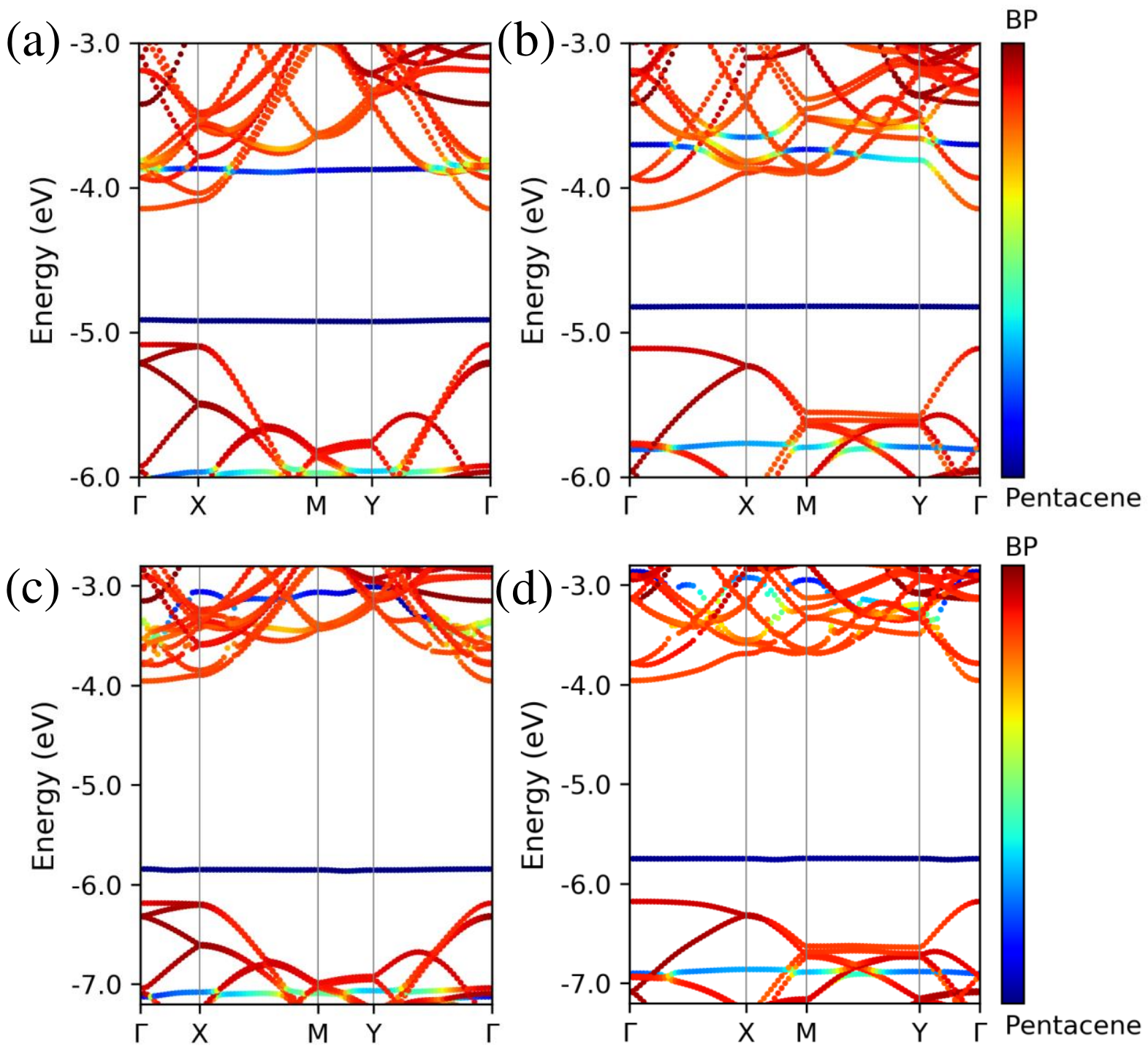}
\caption{Color-coded PBE band structure for (a) the zigzag interface and (b) the armchair interface. Color-coded $GW$ band structure for (c) the zigzag interface and (d) the armchair interface. Fractional coordinates: X=(0.5,0.0,0.0); M=(0.5,0.5,0.0); Y=(0.0,0.5,0.0).}
\label{fig:bands}
\end{figure}

Fig. \ref{fig:bands}(a)(b) show the band structures of the two pentacene:BP interfaces calculated using the PBE functional. To highlight the contributions from each component, we color code the bands that are localized on pentacene (BP) with blue (red), via projections of interface orbitals onto orbitals of the freestanding monolayer pentacene (BP). In the zigzag (armchair) interface, the adsorption of the monolayer pentacene leads to a 0.25 eV (0.18 eV) reduction in the vacuum level compared to that of the pristine monolayer BP, with the latter set to zero in all panels of Fig. \ref{fig:bands}. Both interfaces form a type-II heterostructure, with the highest occupied molecular orbital (HOMO) of pentacene above the valence band maximum (VBM) of BP. In both cases, the pentacene HOMO is dispersionless (in contrast to the herringbone structure that the pentacene typically forms in the bulk crystal\cite{ANPM2009}) and has negligible hybridization with BP orbitals. The lowest unoccupied molecular orbital (LUMO) of pentacene, on the other hand, hybridizes with the BP orbitals in certain regions of the Brillouin zone, due to being close in energy with BP conduction bands.
 
It is well-known that semi-local functionals could not quantitatively describe the interface quasiparticle electronic structure. We then proceed with $G_0W_0$@PBE calculations for an accurate description of the interfacial level alignment. We include 9600 bands (corresponding to 6.0 Ry) in the summation to compute the static Kohn-Sham polarizability in the random-phase approximation. The self-energies are evaluated using the Hybertsen–Louie generalized plasmon-pole model\cite{HL1986} for the frequency dependence of the dielectric function, the semiconductor screening for the treatment of the $q\to0$ limit, the slab Coulomb truncation\cite{SIB2006}, and the static remainder\cite{DSJC2013} to improve convergence, as implemented in the BerkeleyGW package\cite{DSSJ2012}. As a convergence study, we have performed the same $G_0W_0$@PBE calculations for the monolayer BP unit cell, using 800 bands (corresponding to 6.0 Ry) in the calculation of the dielectric function, commensurate with the parameters used for the interface. We found a band gap of 2.13 eV, in good agreement with results from prior $GW$ calculations (2.0 eV\cite{TSLY2014}).

The $GW$ band structures of both interfaces are presented in Fig. \ref{fig:bands}(c)(d). Since we perform $G_0W_0$ calculations, the color code for the bands in Fig. \ref{fig:bands}(c) [(d)] is adopted from that in Fig. \ref{fig:bands}(a) [(b)] and only the eigenvalues are altered. Unlike the PBE results where the interface conduction band minimum (CBM) is localized on the pentacene in some parts of the Brillouin zone, in $GW$ results, the pentacene LUMO is fully embedded in the BP conduction bands across the Brillouin zone. Table \ref{tab:gaps} summarizes the key properties describing the quasiparticle electronic structure of the two orientations of the pentacene:BP interfaces and the freestanding BP and pentacene, evaluated at the $\Gamma$ point. Considering the hybridization between BP and pentacene, we perform two types of $GW$ calculations for the interface, in line with our prior work.\cite{AL2021} The first type is interface $GW$, i.e., $\braket{\phi^{\rm tot}|\Sigma[G^{\rm tot}W^{\rm tot}]|\phi^{\rm tot}}$. Here, $\phi^{\rm tot}$ is an interface orbital that is a resonance arising from one of the following: pentacene HOMO, pentacene LUMO, BP VBM, and BP CBM. The second type is projection $GW$, i.e., $\braket{\phi^{\rm A}|\Sigma[G^{\rm tot}W^{\rm tot}]|\phi^{\rm A}}$. Here, $\phi^{\rm A}$ is one of the aforementioned frontier orbitals that are obtained from standalone calculations of the corresponding species in their freestanding forms. In both cases, $G$ and $W$ are evaluated using the full interface, so that any difference between the two approaches is a reflection of the orbital hybridization (i.e., $\phi^{\rm tot}$ versus $\phi^{\rm A}$). For strongly hybridized orbitals, the difference between these two approaches can be rationalized by considering the off-diagonal matrix elements of the self-energy.\cite{AL2021}

\begin{table}
\caption{$GW$ gaps for the two orientations of the pentacene:BP interfaces, as well as for each individual component in its freestanding form. $E_{\rm g}^{\rm BP}$ ($E_{\rm g}^{\rm mol}$) is the band gap of the BP (monolayer pentacene). $\Delta_{\rm HH}$ ($\Delta_{\rm LL}$) is the relative energy level alignment between the VBM (CBM) of BP and the HOMO (LUMO) of pentacene in the interface. $E_{\rm g}^{\rm tot}$ is the band gap of the interface, between the HOMO of pentacene and the CBM of BP. All values are evaluated at the $\Gamma$ point and in eV.}
\centering
\begin{tabular}{c|c|cc|cc}
\hline\hline
 & Freestanding & \multicolumn{2}{c|}{Zigzag interface} & \multicolumn{2}{c}{Armchair interface} \\
 & & Interface $GW$ & Projection $GW$ & Interface $GW$ & Projection $GW$ \\
\hline
$E_{\rm g}^{\rm BP}$ & 2.20 & 2.23 & 1.95 & 2.22 & 1.95 \\
$E_{\rm g}^{\rm mol}$ & 4.29 & 2.47 & 2.97 & 2.89 & 2.91\\
$\Delta_{\rm HH}$ & $-$ & 0.34 & 0.26 & 0.43 & 0.34 \\
$\Delta_{\rm LL}$ & $-$ & 0.58 & 1.28 & 1.09 & 1.29 \\
$E_{\rm g}^{\rm tot}$ & $-$ & 1.89 & 1.69 & 1.79 & 1.62 \\
\hline\hline
\end{tabular}

\label{tab:gaps}
\end{table}

A few discussions are in place for the results shown in Table \ref{tab:gaps}. First, if we neglect the hybridization and solely consider the dielectric effects of the interface, we find that both orientations of the interface exhibit similar properties, as evidenced by the similarity in the projection $GW$ results for the two orientations of the interface, where the BP gap is reduced by about 0.25 eV and the pentacene gap is reduced by about 1.3 eV compared to their respective freestanding phases. This is consistent with the fact that the adsorption height is similar in both orientations. Second, the difference between interface $GW$ results and projection $GW$ results indicate the strength of hybridization. From Table \ref{tab:gaps} and Fig. \ref{fig:bands}(a)(b), we can clearly see that the zigzag interface features stronger hybridization at the $\Gamma$ point, especially for the pentacene LUMO. This hybridization leads to the difference in $E_{\rm g}^{\rm mol}$ and $\Delta_{\rm LL}$ values computed using the two approaches. Third, we note that Table \ref{tab:gaps} only shows the properties evaluated at the $\Gamma$ point. The $GW$ band structures presented in Fig. \ref{fig:bands}(c)(d) illustrate the band alignment across the Brillouin zone, where the hybridization between pentacene LUMO and BP conduction bands is more pronounced in the armchair interface.

After we discuss the quasiparticle electronic structure, we proceed with BSE calculations of the optical properties. We include 20 valance and 20 conduction bands in the active space in computing the interaction kernel for the interface, 4 valance and 4 conduction bands in the calculation of the freestanding monolayer BP unit cell, and 12 valance and 12 conduction bands in the calculation of the freestanding monolayer pentacene. The BSE Hamiltonian is then interpolated from a coarse $\mathbf{k}$-mesh of 2$\times$4$\times$1 (4$\times$2$\times$1) to a finer $\mathbf{k}$-mesh of 6$\times$12$\times$1 (12$\times$6$\times$1) for the zigzag (armchair) interface before the diagonalization, which corresponds to a fine $\mathbf{k}$-mesh of 36$\times$24$\times$1 for the monolayer BP unit cell. As convergence studies, our parameters for the monolayer BP unit cell yield an optical absorption peak of 1.18 eV (2.80 eV) when the light polarization is along the armchair (zigzag) direction, in good agreement with Ref. \citenum{TSLY2014} where these values are 1.2 eV and 2.8 eV, respectively.

For the heterogeneous interfaces, eigenvectors of the BSE Hamiltonian represent excited states of the entire interface, while it is often useful and intuitive to decompose these interface excited states into different contributions - those localized on the adsorbate, those localized on the substrate, and the charge-transfer ones across the interface - to unravel the interface effects on the excitonic properties of each individual component. Here, we develop a new and general computational scheme to achieve this goal and apply it to the pentacene:BP systems, which we show leads to the assignment of the peaks in the absorption spectra and quantitative insight into the change of the excitonic properties after the formation of the interface. We note that this scheme of computational analysis is applicable to any weakly coupled interfaces (i.e., without bond breaking upon formation of the interface). 

For each excited state of the interface, i.e., an eigenvector of the BSE Hamiltonian of the entire interface, one can write (for simplicity, we suppress the $\mathbf{k}$ indices in the equations)
\begin{equation}
\Psi(\mathbf{r}_{\rm e},\mathbf{r}_{\rm h})=\sum_{v \in {\rm tot}}^{\rm occ.}\sum_{c \in {\rm tot}}^{\rm vir.}A_{vc}\phi_v^*(\mathbf{r}_{\rm h})\phi_c(\mathbf{r}_{\rm e}).
\label{eq:bse}
\end{equation}
Here, $\mathbf{r}_{\rm e}$ ($\mathbf{r}_{\rm h}$) is the coordinate of the electron (hole); $v$ ($c$) runs over all occupied (unoccupied) Kohn-Sham orbitals of the interface, $\phi_v$ ($\phi_c$). $A_{vc}$ is the expansion coefficient. We then perform linear expansions of interface orbitals (``tot'') in terms of the orbitals of each component (the molecule, ``mol'', and the substrate, ``sub'') in its freestanding phase:
\begin{equation}
\begin{split}
\phi_v^*(\mathbf{r}_{\rm h}) & = \sum_{i \in {\rm mol}}^{\rm occ.}C_{vi}^*\phi_i^*(\mathbf{r}_{\rm h})+\sum_{j \in {\rm sub}}^{\rm occ.}C_{vj}^*\phi_j^*(\mathbf{r}_{\rm h});\\
\phi_c(\mathbf{r}_{\rm e}) & = \sum_{a \in {\rm mol}}^{\rm vir.}C_{ca}\phi_a(\mathbf{r}_{\rm e})+\sum_{b \in {\rm sub}}^{\rm vir.}C_{cb}\phi_b(\mathbf{r}_{\rm e}).
\end{split}
\label{eq:exp}
\end{equation}
Here, $i$ and $j$ ($a$ and $b$) run over the occupied (unoccupied) orbitals of the freestanding molecular layer and the freestanding substrate, respectively, and the $C$'s are expansion coefficients. Substituting Eq. \eqref{eq:exp} into Eq. \eqref{eq:bse}, we find
\begin{equation}
\begin{split}
\Psi(\mathbf{r}_{\rm e},\mathbf{r}_{\rm h}) = & \sum_{v \in {\rm tot}}^{\rm occ.}\sum_{c \in {\rm tot}}^{\rm vir.}A_{vc}\sum_{i \in {\rm mol}}^{\rm occ.}\sum_{a \in {\rm mol}}^{\rm vir.}C_{vi}^*C_{ca}\phi_i^*(\mathbf{r}_{\rm h})\phi_a(\mathbf{r}_{\rm e}) \\
+ & \sum_{v \in {\rm tot}}^{\rm occ.}\sum_{c \in {\rm tot}}^{\rm vir.}A_{vc}\sum_{j \in {\rm sub}}^{\rm occ.}\sum_{b \in {\rm sub}}^{\rm vir.}C_{vj}^*C_{cb}\phi_j^*(\mathbf{r}_{\rm h})\phi_b(\mathbf{r}_{\rm e}) \\
+ & \sum_{v \in {\rm tot}}^{\rm occ.}\sum_{c \in {\rm tot}}^{\rm vir.}A_{vc}\sum_{i \in {\rm mol}}^{\rm occ.}\sum_{b \in {\rm sub}}^{\rm vir.}C_{vi}^*C_{cb}\phi_i^*(\mathbf{r}_{\rm h})\phi_b(\mathbf{r}_{\rm e}) \\
+ & \sum_{v \in {\rm tot}}^{\rm occ.}\sum_{c \in {\rm tot}}^{\rm vir.}A_{vc}\sum_{j \in {\rm sub}}^{\rm occ.}\sum_{a \in {\rm mol}}^{\rm vir.}C_{vj}^*C_{ca}\phi_j^*(\mathbf{r}_{\rm h})\phi_a(\mathbf{r}_{\rm e}). \\
\end{split}
\label{eq:decomp}
\end{equation}
The term in the first row of Eq. \eqref{eq:decomp} has both the hole $\mathbf{r}_{\rm h}$ and the electron $\mathbf{r}_{\rm e}$ localized on the molecule, therefore describes an excited state localized on the molecule. Similarly, the term in the second row describes an excited state localized on the substrate. The term in the third (fourth) row has the hole $\mathbf{r}_{\rm h}$ localized on the molecule (substrate) and the electron $\mathbf{r}_{\rm e}$ localized on the substrate (molecule), therefore describes a charge-transfer excitation from the molecule (substrate) to the substrate (molecule).

\begin{figure}[h]
\centering
\includegraphics[width=4in]{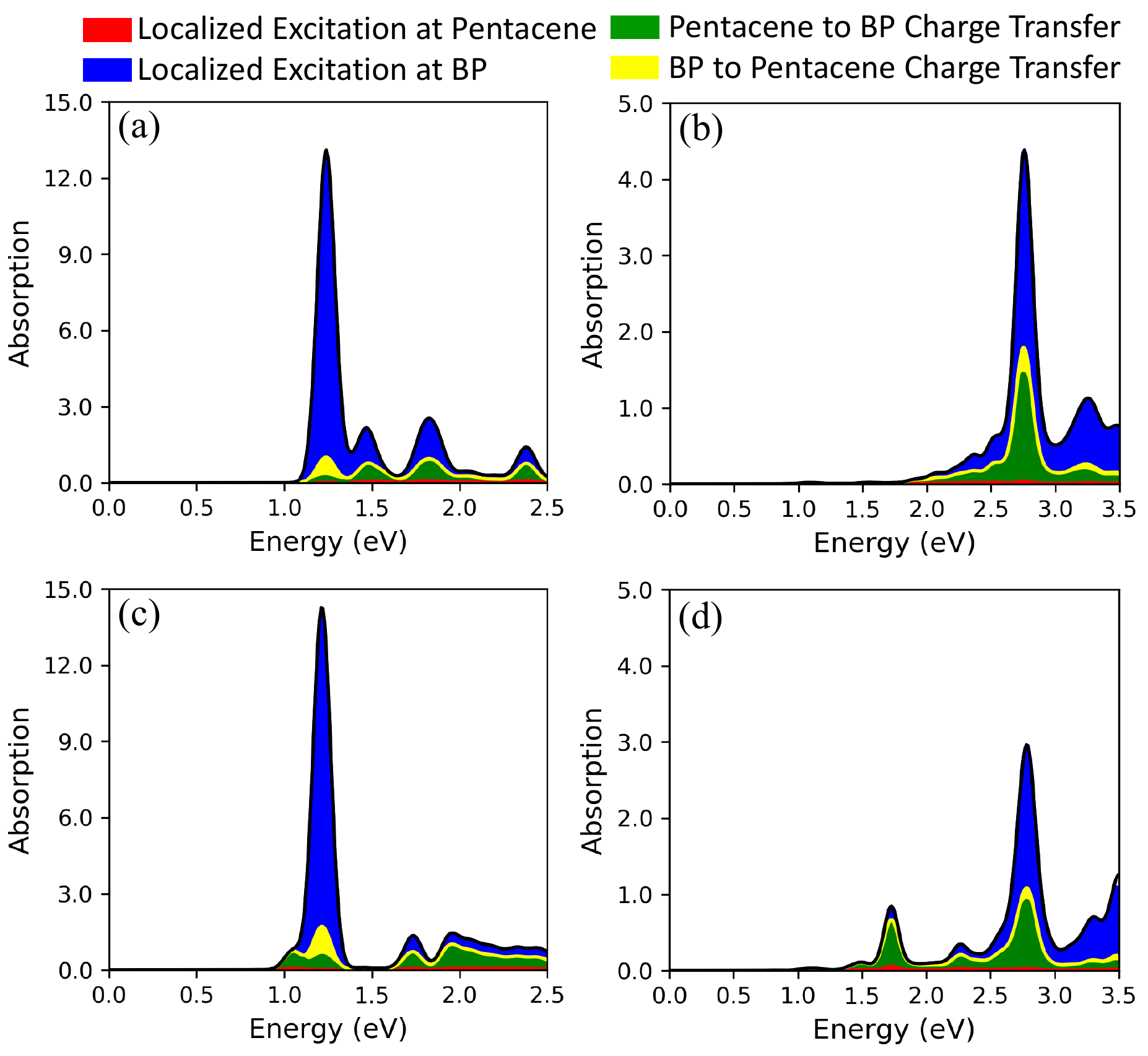}
\caption{Decomposition of the optical absorption spectra of pentacene:BP interfaces. (a) and (b) [(c) and (d)] are for the zigzag (armchair) interface with the light polarization along the armchair and zigzag direction, respectively.}
\label{fig:bse}
\end{figure}

In Fig. \ref{fig:bse}, we show the BSE optical absorption spectra for the four cases, namely the zigzag interface [(a) and (b)] and the armchair interface [(c) and (d)], respectively, with the light polarization along either the armchair [(a) and (c)] or the zigzag direction [(b) and (d)] of monolayer BP, respectively. Furthermore, we decompose the absorption spectra based on the different contributions to all excited states that collectively give rise to the absorption spectra, as calculated using Eq. \eqref{eq:decomp}. We first notice that the optical anisotropy of both interfaces largely follows that of the monolayer BP\cite{TSLY2014}, i.e., the first major peak is around 1.2 eV when the light polarization is along the armchair direction [(a) and (c)], much lower in energy than the cases when the light polarization is along the zigzag direction [(b) and (d), about 2.8 eV]. This intrinsic BP anisotropy is already well understood in terms of dipole selection rules \cite{LA2014,TFY2015}. In all panels, the absorption magnitude stemming from pentacene-localized excitations is negligible compared to other contributions.

Notably, the major difference between the optical absorption spectra of both interfaces and that of the pristine monolayer BP is the low-lying peak around 1.7 eV in Fig. \ref{fig:bse}(d), which has been assigned as pentacene-to-BP charge transfer. A detailed investigation of this peak indicates that it is primarily the transition from the interface VBM (arising from the pentacene HOMO) to the interface CBM (arising from monolayer BP CBM). The same transition is also present in Fig. \ref{fig:bse}(a), but overlaps with the secondary and tertiary peaks arising from BP-localized transitions, making it less prominent than the situation presented in Fig. \ref{fig:bse}(d). In Fig. \ref{fig:bse}(b), this interface VBM-CBM transition is also at similar energy, but with a nearly negligible oscillator strength. This is because in Fig. \ref{fig:bse}(b), the light polarization is along the zigzag direction in the zigzag interface, which coincides with the long axis of monolayer pentacene, which is not the favorable direction for the light polarization as far as the optical absorption of monolayer pentacene is concerned.\cite{CGR2012,SWWC2014} In Fig. \ref{fig:bse}(c), this charge-transfer excitation is manifested in two regions: between 1.0 eV and 1.2 eV, and around 1.7 eV. In both regions, the oscillator strengths of the charge-transfer excitations are overshadowed by those of the BP-localized excitations, making their actual detection and application more difficult than their counterparts in Fig. \ref{fig:bse}(d). Overall, we conclude that the 1.7 eV charge-transfer peak seen in Fig. \ref{fig:bse}(d) is characteristic of this interface, as it appears at an energy range where the BP-localized optical absorption is negligible. Similar features might be observed for other anisotropic interfaces.

\begin{figure}[h]
\centering
\includegraphics[width=4in]{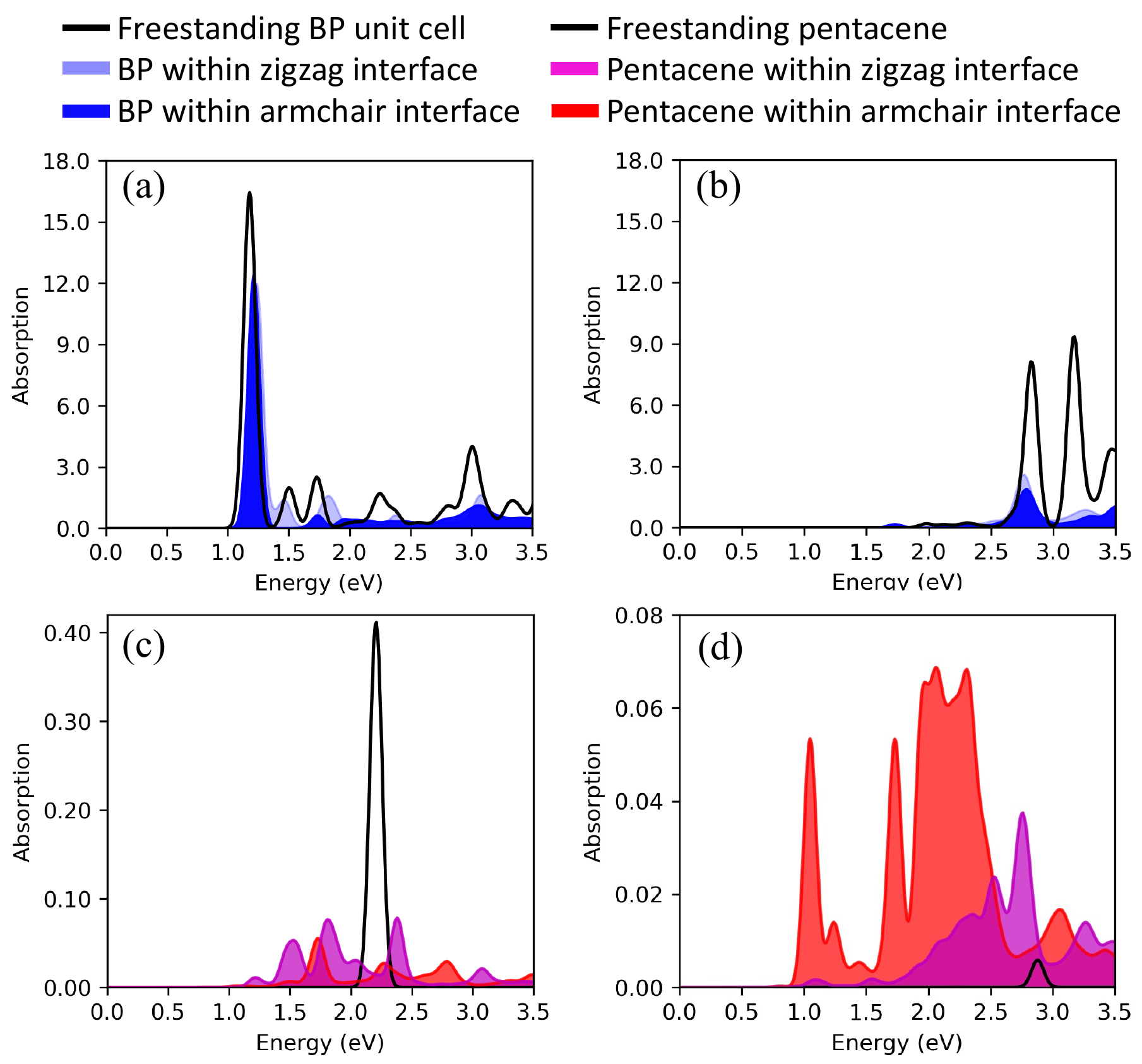}
\caption{Comparison of optical absorption spectra for monolayer BP when the light polarization is along the (a) armchair and (b) zigzag direction. In both (a) and (b), black is for the monolayer BP unit cell in its freestanding phase. Light blue (dark blue) is for BP-localized excited states within the zigzag (armchair) interface. Comparison of optical absorption spectra for monolayer pentacene when the light polarization is along the (c) short and (d) long axis of pentacene. In both (c) and (d), black is for the monolayer pentacene in its freestanding phase. Magenta (red) is for pentacene-localized excited states within the zigzag (armchair) interface.}
\label{fig:compare}
\end{figure}

Lastly, the new exciton decomposition scheme we developed here allows us to quantitatively address the two questions we asked at the beginning of this paper, i.e., how the anisotropies of BP and the adsorbate are modulated by the formation of the interface. Fig. \ref{fig:compare}(a) [Fig. \ref{fig:compare}(b)] compares the optical absorption of a freestanding monolayer BP unit cell with the absorption spectra arising from BP-localized excited states (as obtained from the decomposition analysis results shown in Fig. \ref{fig:bse}) in the pentacene:BP interfaces, when the polarization of the light is along the armchair (zigzag) direction. In both Fig. \ref{fig:compare}(a) and (b), the black line is the pristine BP absorption, and the light blue (dark blue) shaded area denotes the case of zigzag (armchair) interface. One can see that in Fig. \ref{fig:compare}(a) [Fig. \ref{fig:compare}(b)], the lowest interface peak is slightly blueshifted (redshifted) by about 0.1 eV, for both orientations of the interface. The positions of higher excitations are also modulated by the adsorption of pentacene.

A similar comparison is presented in Fig. \ref{fig:compare}(c)(d), where Fig. \ref{fig:compare}(c) [Fig. \ref{fig:compare}(d)] is the optical absorption of the monolayer pentacene when the light polarization is along the short (long) axis of the molecule. Note that in the zigzag (armchair) interface, the short axis of pentacene coincides with the armchair (zigzag) direction of BP. The black line is the pristine monolayer pentacene absorption corresponding to the HOMO-LUMO transition, consistent with previous results.\cite{CGR2012,SWWC2014} Magenta (red) shaded area denotes the optical absorption arising from pentacene-localized excited states (as obtained from Fig. \ref{fig:bse}) in the zigzag (armchair) pentacene:BP interface. One can see that the formation of the interface introduces a redshift to the pentacene-localized optical excitation. The complex envelope of the decomposed interface absorption spectra is a result of orbital hybridization upon the formation of the interface.

In summary, we have studied the anisotropy of the quasiparticle and optical properties of pentacene:BP interfaces in two orientations, where the monolayer pentacene is aligned along the zigzag or the armchair direction of monolayer BP. To assign peaks in the absorption spectra and to obtain a quantitative insight into the interface effects in modulating the optical properties, we have developed a new and general computational analysis tool, with which we decomposed the interface excited states into those localized on each individual component plus charge-transfer ones. Our analysis of the optical properties reveals a distinct charge-transfer absorption peak, for the armchair interface with light polarization along the zigzag direction. Our study of this prototypical system leads to a detailed understanding of the interaction between monolayer BP and molecular adsorbates, which paves the way for the future development of surface passivation of anisotropic layered materials.

\section{Acknowledgements}
Z.-F.L. acknowledges an NSF CAREER Award, DMR-2044552. This work is also supported by the Faculty Competition for Postdoctoral Fellows award from the Office of the Vice President for Research at Wayne State University. This work uses computational resources at the Center for Functional Nanomaterials (CFN), which is a U.S. Department of Energy Office of Science User Facility, at Brookhaven National Laboratory under Contract No. DE-SC0012704. Additional large-scale benchmark studies are performed using resources of the National Energy Research Scientific Computing Center (NERSC), a U.S. Department of Energy Office of Science User Facility located at Lawrence Berkeley National Laboratory, operated under Contract No. DE-AC02-05CH11231 through NERSC award BES-ERCAP0020328, as well as the Extreme Science and Engineering Discovery Environment (XSEDE), which is supported by National Science Foundation grant number ACI-1548562 through allocation PHY220043.

\bibliography{ref.bib}

\providecommand{\latin}[1]{#1}
\makeatletter
\providecommand{\doi}
  {\begingroup\let\do\@makeother\dospecials
  \catcode`\{=1 \catcode`\}=2 \doi@aux}
\providecommand{\doi@aux}[1]{\endgroup\texttt{#1}}
\makeatother
\providecommand*\mcitethebibliography{\thebibliography}
\csname @ifundefined\endcsname{endmcitethebibliography}
  {\let\endmcitethebibliography\endthebibliography}{}
\begin{mcitethebibliography}{59}
\providecommand*\natexlab[1]{#1}
\providecommand*\mciteSetBstSublistMode[1]{}
\providecommand*\mciteSetBstMaxWidthForm[2]{}
\providecommand*\mciteBstWouldAddEndPuncttrue
  {\def\EndOfBibitem{\unskip.}}
\providecommand*\mciteBstWouldAddEndPunctfalse
  {\let\EndOfBibitem\relax}
\providecommand*\mciteSetBstMidEndSepPunct[3]{}
\providecommand*\mciteSetBstSublistLabelBeginEnd[3]{}
\providecommand*\EndOfBibitem{}
\mciteSetBstSublistMode{f}
\mciteSetBstMaxWidthForm{subitem}{(\alph{mcitesubitemcount})}
\mciteSetBstSublistLabelBeginEnd
  {\mcitemaxwidthsubitemform\space}
  {\relax}
  {\relax}

\bibitem[Xia \latin{et~al.}(2014)Xia, Wang, and Jia]{XWJ2014}
Xia,~F.; Wang,~H.; Jia,~Y. Rediscovering black phosphorus as an anisotropic
  layered material for optoelectronics and electronics. \emph{Nat. Commun.}
  \textbf{2014}, \emph{5}, 4458\relax
\mciteBstWouldAddEndPuncttrue
\mciteSetBstMidEndSepPunct{\mcitedefaultmidpunct}
{\mcitedefaultendpunct}{\mcitedefaultseppunct}\relax
\EndOfBibitem
\bibitem[Ling \latin{et~al.}(2015)Ling, Wang, Huang, Xia, and
  Dresselhaus]{LWHX2014}
Ling,~X.; Wang,~H.; Huang,~S.; Xia,~F.; Dresselhaus,~M.~S. The renaissance of
  black phosphorus. \emph{Proc. Natl. Acad. Sci.} \textbf{2015}, \emph{112},
  4523--4530\relax
\mciteBstWouldAddEndPuncttrue
\mciteSetBstMidEndSepPunct{\mcitedefaultmidpunct}
{\mcitedefaultendpunct}{\mcitedefaultseppunct}\relax
\EndOfBibitem
\bibitem[Carvalho \latin{et~al.}(2016)Carvalho, Wang, Zhu, Rodin, Su, and
  Castro~Neto]{CWZR2016}
Carvalho,~A.; Wang,~M.; Zhu,~X.; Rodin,~A.~S.; Su,~H.; Castro~Neto,~A.~H.
  Phosphorene: from theory to applications. \emph{Nat. Rev. Mater.}
  \textbf{2016}, \emph{1}, 16061\relax
\mciteBstWouldAddEndPuncttrue
\mciteSetBstMidEndSepPunct{\mcitedefaultmidpunct}
{\mcitedefaultendpunct}{\mcitedefaultseppunct}\relax
\EndOfBibitem
\bibitem[Mao \latin{et~al.}(2016)Mao, Tang, Xie, Wu, Han, Lin, Deng, Ji, Xu,
  Liu, \latin{et~al.} others]{MTXW2016}
Mao,~N.; Tang,~J.; Xie,~L.; Wu,~J.; Han,~B.; Lin,~J.; Deng,~S.; Ji,~W.; Xu,~H.;
  Liu,~K. \latin{et~al.}  Optical anisotropy of black phosphorus in the visible
  regime. \emph{J. Am. Chem. Soc.} \textbf{2016}, \emph{138}, 300--305\relax
\mciteBstWouldAddEndPuncttrue
\mciteSetBstMidEndSepPunct{\mcitedefaultmidpunct}
{\mcitedefaultendpunct}{\mcitedefaultseppunct}\relax
\EndOfBibitem
\bibitem[Tran \latin{et~al.}(2014)Tran, Soklaski, Liang, and Yang]{TSLY2014}
Tran,~V.; Soklaski,~R.; Liang,~Y.; Yang,~L. Layer-controlled band gap and
  anisotropic excitons in few-layer black phosphorus. \emph{Phys. Rev. B}
  \textbf{2014}, \emph{89}, 235319\relax
\mciteBstWouldAddEndPuncttrue
\mciteSetBstMidEndSepPunct{\mcitedefaultmidpunct}
{\mcitedefaultendpunct}{\mcitedefaultseppunct}\relax
\EndOfBibitem
\bibitem[Low \latin{et~al.}(2014)Low, Rodin, Carvalho, Jiang, Wang, Xia, and
  Neto]{LRCJ2014}
Low,~T.; Rodin,~A.; Carvalho,~A.; Jiang,~Y.; Wang,~H.; Xia,~F.; Neto,~A.~C.
  Tunable optical properties of multilayer black phosphorus thin films.
  \emph{Phys. Rev. B} \textbf{2014}, \emph{90}, 075434\relax
\mciteBstWouldAddEndPuncttrue
\mciteSetBstMidEndSepPunct{\mcitedefaultmidpunct}
{\mcitedefaultendpunct}{\mcitedefaultseppunct}\relax
\EndOfBibitem
\bibitem[Wang \latin{et~al.}(2015)Wang, Jones, Seyler, Tran, Jia, Zhao, Wang,
  Yang, Xu, and Xia]{WJST2015}
Wang,~X.; Jones,~A.~M.; Seyler,~K.~L.; Tran,~V.; Jia,~Y.; Zhao,~H.; Wang,~H.;
  Yang,~L.; Xu,~X.; Xia,~F. Highly anisotropic and robust excitons in monolayer
  black phosphorus. \emph{Nat. Nanotechnol.} \textbf{2015}, \emph{10},
  517--521\relax
\mciteBstWouldAddEndPuncttrue
\mciteSetBstMidEndSepPunct{\mcitedefaultmidpunct}
{\mcitedefaultendpunct}{\mcitedefaultseppunct}\relax
\EndOfBibitem
\bibitem[Lan \latin{et~al.}(2016)Lan, Rodrigues, Kang, and Cai]{LRKC2016}
Lan,~S.; Rodrigues,~S.; Kang,~L.; Cai,~W. Visualizing optical phase anisotropy
  in black phosphorus. \emph{ACS Photonics} \textbf{2016}, \emph{3},
  1176--1181\relax
\mciteBstWouldAddEndPuncttrue
\mciteSetBstMidEndSepPunct{\mcitedefaultmidpunct}
{\mcitedefaultendpunct}{\mcitedefaultseppunct}\relax
\EndOfBibitem
\bibitem[Fei \latin{et~al.}(2014)Fei, Faghaninia, Soklaski, Yan, Lo, and
  Yang]{FFSY2014}
Fei,~R.; Faghaninia,~A.; Soklaski,~R.; Yan,~J.-A.; Lo,~C.; Yang,~L. Enhanced
  thermoelectric efficiency via orthogonal electrical and thermal conductances
  in phosphorene. \emph{Nano Lett.} \textbf{2014}, \emph{14}, 6393--6399\relax
\mciteBstWouldAddEndPuncttrue
\mciteSetBstMidEndSepPunct{\mcitedefaultmidpunct}
{\mcitedefaultendpunct}{\mcitedefaultseppunct}\relax
\EndOfBibitem
\bibitem[Qiao \latin{et~al.}(2014)Qiao, Kong, Hu, Yang, and Ji]{QKHY2014}
Qiao,~J.; Kong,~X.; Hu,~Z.-X.; Yang,~F.; Ji,~W. High-mobility transport
  anisotropy and linear dichroism in few-layer black phosphorus. \emph{Nat.
  Commun.} \textbf{2014}, \emph{5}, 4475\relax
\mciteBstWouldAddEndPuncttrue
\mciteSetBstMidEndSepPunct{\mcitedefaultmidpunct}
{\mcitedefaultendpunct}{\mcitedefaultseppunct}\relax
\EndOfBibitem
\bibitem[He \latin{et~al.}(2015)He, He, Wang, Cui, Bellus, Chiu, and
  Zhao]{HHWC2015}
He,~J.; He,~D.; Wang,~Y.; Cui,~Q.; Bellus,~M.~Z.; Chiu,~H.-Y.; Zhao,~H.
  Exceptional and anisotropic transport properties of photocarriers in black
  phosphorus. \emph{ACS Nano} \textbf{2015}, \emph{9}, 6436--6442\relax
\mciteBstWouldAddEndPuncttrue
\mciteSetBstMidEndSepPunct{\mcitedefaultmidpunct}
{\mcitedefaultendpunct}{\mcitedefaultseppunct}\relax
\EndOfBibitem
\bibitem[Luo \latin{et~al.}(2015)Luo, Maassen, Deng, Du, Garrelts, Lundstrom,
  Ye, and Xu]{LMDD2015}
Luo,~Z.; Maassen,~J.; Deng,~Y.; Du,~Y.; Garrelts,~R.~P.; Lundstrom,~M.~S.;
  Ye,~P.~D.; Xu,~X. Anisotropic in-plane thermal conductivity observed in
  few-layer black phosphorus. \emph{Nat. Commun.} \textbf{2015}, \emph{6},
  8572\relax
\mciteBstWouldAddEndPuncttrue
\mciteSetBstMidEndSepPunct{\mcitedefaultmidpunct}
{\mcitedefaultendpunct}{\mcitedefaultseppunct}\relax
\EndOfBibitem
\bibitem[Lee \latin{et~al.}(2015)Lee, Yang, Suh, Yang, Lee, Li, Sung~Choe,
  Suslu, Chen, Ko, \latin{et~al.} others]{LYSY2015}
Lee,~S.; Yang,~F.; Suh,~J.; Yang,~S.; Lee,~Y.; Li,~G.; Sung~Choe,~H.;
  Suslu,~A.; Chen,~Y.; Ko,~C. \latin{et~al.}  Anisotropic in-plane thermal
  conductivity of black phosphorus nanoribbons at temperatures higher than 100
  K. \emph{Nat. Commun.} \textbf{2015}, \emph{6}, 8573\relax
\mciteBstWouldAddEndPuncttrue
\mciteSetBstMidEndSepPunct{\mcitedefaultmidpunct}
{\mcitedefaultendpunct}{\mcitedefaultseppunct}\relax
\EndOfBibitem
\bibitem[Low \latin{et~al.}(2014)Low, Rold\'an, Wang, Xia, Avouris, Moreno, and
  Guinea]{LRWX2014}
Low,~T.; Rold\'an,~R.; Wang,~H.; Xia,~F.; Avouris,~P.; Moreno,~L.~M.;
  Guinea,~F. Plasmons and Screening in Monolayer and Multilayer Black
  Phosphorus. \emph{Phys. Rev. Lett.} \textbf{2014}, \emph{113}, 106802\relax
\mciteBstWouldAddEndPuncttrue
\mciteSetBstMidEndSepPunct{\mcitedefaultmidpunct}
{\mcitedefaultendpunct}{\mcitedefaultseppunct}\relax
\EndOfBibitem
\bibitem[Liu and Aydin(2016)Liu, and Aydin]{LA2016}
Liu,~Z.; Aydin,~K. Localized surface plasmons in nanostructured monolayer black
  phosphorus. \emph{Nano Lett.} \textbf{2016}, \emph{16}, 3457--3462\relax
\mciteBstWouldAddEndPuncttrue
\mciteSetBstMidEndSepPunct{\mcitedefaultmidpunct}
{\mcitedefaultendpunct}{\mcitedefaultseppunct}\relax
\EndOfBibitem
\bibitem[Correas-Serrano \latin{et~al.}(2017)Correas-Serrano, Al{\`u}, and
  Gomez-Diaz]{CAD2017}
Correas-Serrano,~D.; Al{\`u},~A.; Gomez-Diaz,~J.~S. Plasmon canalization and
  tunneling over anisotropic metasurfaces. \emph{Phys. Rev. B} \textbf{2017},
  \emph{96}, 075436\relax
\mciteBstWouldAddEndPuncttrue
\mciteSetBstMidEndSepPunct{\mcitedefaultmidpunct}
{\mcitedefaultendpunct}{\mcitedefaultseppunct}\relax
\EndOfBibitem
\bibitem[Ling \latin{et~al.}(2016)Ling, Huang, Hasdeo, Liang, Parkin, Tatsumi,
  Nugraha, Puretzky, Das, Sumpter, \latin{et~al.} others]{LHHL2016}
Ling,~X.; Huang,~S.; Hasdeo,~E.~H.; Liang,~L.; Parkin,~W.~M.; Tatsumi,~Y.;
  Nugraha,~A.~R.; Puretzky,~A.~A.; Das,~P.~M.; Sumpter,~B.~G. \latin{et~al.}
  Anisotropic electron-photon and electron-phonon interactions in black
  phosphorus. \emph{Nano Lett.} \textbf{2016}, \emph{16}, 2260--2267\relax
\mciteBstWouldAddEndPuncttrue
\mciteSetBstMidEndSepPunct{\mcitedefaultmidpunct}
{\mcitedefaultendpunct}{\mcitedefaultseppunct}\relax
\EndOfBibitem
\bibitem[Liao \latin{et~al.}(2015)Liao, Zhou, Qiu, Dresselhaus, and
  Chen]{LZQD2015}
Liao,~B.; Zhou,~J.; Qiu,~B.; Dresselhaus,~M.~S.; Chen,~G. Ab initio study of
  electron-phonon interaction in phosphorene. \emph{Phys. Rev. B}
  \textbf{2015}, \emph{91}, 235419\relax
\mciteBstWouldAddEndPuncttrue
\mciteSetBstMidEndSepPunct{\mcitedefaultmidpunct}
{\mcitedefaultendpunct}{\mcitedefaultseppunct}\relax
\EndOfBibitem
\bibitem[Yuan \latin{et~al.}(2015)Yuan, Liu, Afshinmanesh, Li, Xu, Sun, Lian,
  Curto, Ye, Hikita, \latin{et~al.} others]{YLAL2015}
Yuan,~H.; Liu,~X.; Afshinmanesh,~F.; Li,~W.; Xu,~G.; Sun,~J.; Lian,~B.;
  Curto,~A.~G.; Ye,~G.; Hikita,~Y. \latin{et~al.}  Polarization-sensitive
  broadband photodetector using a black phosphorus vertical p--n junction.
  \emph{Nat. Nanotechnol.} \textbf{2015}, \emph{10}, 707--713\relax
\mciteBstWouldAddEndPuncttrue
\mciteSetBstMidEndSepPunct{\mcitedefaultmidpunct}
{\mcitedefaultendpunct}{\mcitedefaultseppunct}\relax
\EndOfBibitem
\bibitem[Kou \latin{et~al.}(2014)Kou, Frauenheim, and Chen]{KFC2014}
Kou,~L.; Frauenheim,~T.; Chen,~C. Phosphorene as a superior gas sensor:
  selective adsorption and distinct I--V response. \emph{J. Phys. Chem. Lett.}
  \textbf{2014}, \emph{5}, 2675--2681\relax
\mciteBstWouldAddEndPuncttrue
\mciteSetBstMidEndSepPunct{\mcitedefaultmidpunct}
{\mcitedefaultendpunct}{\mcitedefaultseppunct}\relax
\EndOfBibitem
\bibitem[Cai \latin{et~al.}(2014)Cai, Zhang, and Zhang]{CZZ2014}
Cai,~Y.; Zhang,~G.; Zhang,~Y.-W. Layer-dependent band alignment and work
  function of few-layer phosphorene. \emph{Sci. Rep.} \textbf{2014}, \emph{4},
  6677\relax
\mciteBstWouldAddEndPuncttrue
\mciteSetBstMidEndSepPunct{\mcitedefaultmidpunct}
{\mcitedefaultendpunct}{\mcitedefaultseppunct}\relax
\EndOfBibitem
\bibitem[Li \latin{et~al.}(2014)Li, Yu, Ye, Ge, Ou, Wu, Feng, Chen, and
  Zhang]{LYYG2014}
Li,~L.; Yu,~Y.; Ye,~G.~J.; Ge,~Q.; Ou,~X.; Wu,~H.; Feng,~D.; Chen,~X.~H.;
  Zhang,~Y. Black phosphorus field-effect transistors. \emph{Nat. Nanotechnol.}
  \textbf{2014}, \emph{9}, 372--377\relax
\mciteBstWouldAddEndPuncttrue
\mciteSetBstMidEndSepPunct{\mcitedefaultmidpunct}
{\mcitedefaultendpunct}{\mcitedefaultseppunct}\relax
\EndOfBibitem
\bibitem[Liu \latin{et~al.}(2014)Liu, Neal, Zhu, Luo, Xu, Tom{\'a}nek, and
  Ye]{LNZL2014}
Liu,~H.; Neal,~A.~T.; Zhu,~Z.; Luo,~Z.; Xu,~X.; Tom{\'a}nek,~D.; Ye,~P.~D.
  Phosphorene: an unexplored 2D semiconductor with a high hole mobility.
  \emph{ACS Nano} \textbf{2014}, \emph{8}, 4033--4041\relax
\mciteBstWouldAddEndPuncttrue
\mciteSetBstMidEndSepPunct{\mcitedefaultmidpunct}
{\mcitedefaultendpunct}{\mcitedefaultseppunct}\relax
\EndOfBibitem
\bibitem[Guo \latin{et~al.}(2015)Guo, Zhang, Lu, Wang, Tang, Shao, Sun, Xie,
  Wang, Yu, \latin{et~al.} others]{GZLW2015}
Guo,~Z.; Zhang,~H.; Lu,~S.; Wang,~Z.; Tang,~S.; Shao,~J.; Sun,~Z.; Xie,~H.;
  Wang,~H.; Yu,~X.-F. \latin{et~al.}  From black phosphorus to phosphorene:
  basic solvent exfoliation, evolution of Raman scattering, and applications to
  ultrafast photonics. \emph{Adv. Funct. Mater.} \textbf{2015}, \emph{25},
  6996--7002\relax
\mciteBstWouldAddEndPuncttrue
\mciteSetBstMidEndSepPunct{\mcitedefaultmidpunct}
{\mcitedefaultendpunct}{\mcitedefaultseppunct}\relax
\EndOfBibitem
\bibitem[Xia \latin{et~al.}(2014)Xia, Wang, Xiao, Dubey, and
  Ramasubramaniam]{XWXD2014}
Xia,~F.; Wang,~H.; Xiao,~D.; Dubey,~M.; Ramasubramaniam,~A. Two-dimensional
  material nanophotonics. \emph{Nat. Photonics} \textbf{2014}, \emph{8},
  899--907\relax
\mciteBstWouldAddEndPuncttrue
\mciteSetBstMidEndSepPunct{\mcitedefaultmidpunct}
{\mcitedefaultendpunct}{\mcitedefaultseppunct}\relax
\EndOfBibitem
\bibitem[Castellanos-Gomez \latin{et~al.}(2014)Castellanos-Gomez, Vicarelli,
  Prada, Island, Narasimha-Acharya, Blanter, Groenendijk, Buscema, Steele,
  Alvarez, \latin{et~al.} others]{CVPI2014}
Castellanos-Gomez,~A.; Vicarelli,~L.; Prada,~E.; Island,~J.~O.;
  Narasimha-Acharya,~K.; Blanter,~S.~I.; Groenendijk,~D.~J.; Buscema,~M.;
  Steele,~G.~A.; Alvarez,~J. \latin{et~al.}  Isolation and characterization of
  few-layer black phosphorus. \emph{2D Mater.} \textbf{2014}, \emph{1},
  025001\relax
\mciteBstWouldAddEndPuncttrue
\mciteSetBstMidEndSepPunct{\mcitedefaultmidpunct}
{\mcitedefaultendpunct}{\mcitedefaultseppunct}\relax
\EndOfBibitem
\bibitem[Island \latin{et~al.}(2015)Island, Steele, van~der Zant, and
  Castellanos-Gomez]{ISZC2015}
Island,~J.~O.; Steele,~G.~A.; van~der Zant,~H.~S.; Castellanos-Gomez,~A.
  Environmental instability of few-layer black phosphorus. \emph{2D Mater.}
  \textbf{2015}, \emph{2}, 011002\relax
\mciteBstWouldAddEndPuncttrue
\mciteSetBstMidEndSepPunct{\mcitedefaultmidpunct}
{\mcitedefaultendpunct}{\mcitedefaultseppunct}\relax
\EndOfBibitem
\bibitem[Favron \latin{et~al.}(2015)Favron, Gaufr{\`e}s, Fossard,
  Phaneuf-L’Heureux, Tang, L{\'e}vesque, Loiseau, Leonelli, Francoeur, and
  Martel]{FGFP2015}
Favron,~A.; Gaufr{\`e}s,~E.; Fossard,~F.; Phaneuf-L’Heureux,~A.-L.;
  Tang,~N.~Y.; L{\'e}vesque,~P.~L.; Loiseau,~A.; Leonelli,~R.; Francoeur,~S.;
  Martel,~R. Photooxidation and quantum confinement effects in exfoliated black
  phosphorus. \emph{Nat. Mater.} \textbf{2015}, \emph{14}, 826--832\relax
\mciteBstWouldAddEndPuncttrue
\mciteSetBstMidEndSepPunct{\mcitedefaultmidpunct}
{\mcitedefaultendpunct}{\mcitedefaultseppunct}\relax
\EndOfBibitem
\bibitem[Wood \latin{et~al.}(2014)Wood, Wells, Jariwala, Chen, Cho, Sangwan,
  Liu, Lauhon, Marks, and Hersam]{WWJC2014}
Wood,~J.~D.; Wells,~S.~A.; Jariwala,~D.; Chen,~K.-S.; Cho,~E.; Sangwan,~V.~K.;
  Liu,~X.; Lauhon,~L.~J.; Marks,~T.~J.; Hersam,~M.~C. Effective passivation of
  exfoliated black phosphorus transistors against ambient degradation.
  \emph{Nano Lett.} \textbf{2014}, \emph{14}, 6964--6970\relax
\mciteBstWouldAddEndPuncttrue
\mciteSetBstMidEndSepPunct{\mcitedefaultmidpunct}
{\mcitedefaultendpunct}{\mcitedefaultseppunct}\relax
\EndOfBibitem
\bibitem[Avsar \latin{et~al.}(2015)Avsar, Vera-Marun, Tan, Watanabe, Taniguchi,
  Castro~Neto, and Ozyilmaz]{AVTW2015}
Avsar,~A.; Vera-Marun,~I.~J.; Tan,~J.~Y.; Watanabe,~K.; Taniguchi,~T.;
  Castro~Neto,~A.~H.; Ozyilmaz,~B. Air-stable transport in graphene-contacted,
  fully encapsulated ultrathin black phosphorus-based field-effect transistors.
  \emph{ACS Nano} \textbf{2015}, \emph{9}, 4138--4145\relax
\mciteBstWouldAddEndPuncttrue
\mciteSetBstMidEndSepPunct{\mcitedefaultmidpunct}
{\mcitedefaultendpunct}{\mcitedefaultseppunct}\relax
\EndOfBibitem
\bibitem[Liu \latin{et~al.}(2017)Liu, Shivananju, Wang, Zhang, Yu, Xiao, Sun,
  Ma, Mu, Lin, \latin{et~al.} others]{LSWZ2017}
Liu,~Y.; Shivananju,~B.~N.; Wang,~Y.; Zhang,~Y.; Yu,~W.; Xiao,~S.; Sun,~T.;
  Ma,~W.; Mu,~H.; Lin,~S. \latin{et~al.}  Highly efficient and air-stable
  infrared photodetector based on 2D layered graphene--black phosphorus
  heterostructure. \emph{ACS Appl. Mater. Interfaces} \textbf{2017}, \emph{9},
  36137--36145\relax
\mciteBstWouldAddEndPuncttrue
\mciteSetBstMidEndSepPunct{\mcitedefaultmidpunct}
{\mcitedefaultendpunct}{\mcitedefaultseppunct}\relax
\EndOfBibitem
\bibitem[Doganov \latin{et~al.}(2015)Doganov, O’farrell, Koenig, Yeo,
  Ziletti, Carvalho, Campbell, Coker, Watanabe, Taniguchi, \latin{et~al.}
  others]{DOKY2015}
Doganov,~R.~A.; O’farrell,~E.~C.; Koenig,~S.~P.; Yeo,~Y.; Ziletti,~A.;
  Carvalho,~A.; Campbell,~D.~K.; Coker,~D.~F.; Watanabe,~K.; Taniguchi,~T.
  \latin{et~al.}  Transport properties of pristine few-layer black phosphorus
  by van der Waals passivation in an inert atmosphere. \emph{Nat. Commun.}
  \textbf{2015}, \emph{6}, 6647\relax
\mciteBstWouldAddEndPuncttrue
\mciteSetBstMidEndSepPunct{\mcitedefaultmidpunct}
{\mcitedefaultendpunct}{\mcitedefaultseppunct}\relax
\EndOfBibitem
\bibitem[Wang \latin{et~al.}(2017)Wang, Niu, Liu, Wang, Wei, Liu, Xie, and
  Gao]{WNLW2017}
Wang,~C.; Niu,~D.; Liu,~B.; Wang,~S.; Wei,~X.; Liu,~Y.; Xie,~H.; Gao,~Y. Charge
  transfer at the PTCDA/black phosphorus interface. \emph{J. Phys. Chem. C}
  \textbf{2017}, \emph{121}, 18084--18094\relax
\mciteBstWouldAddEndPuncttrue
\mciteSetBstMidEndSepPunct{\mcitedefaultmidpunct}
{\mcitedefaultendpunct}{\mcitedefaultseppunct}\relax
\EndOfBibitem
\bibitem[Yue \latin{et~al.}(2016)Yue, Lee, Jang, Choi, Nam, Jung, and
  Yoo]{YLJC2016}
Yue,~D.; Lee,~D.; Jang,~Y.~D.; Choi,~M.~S.; Nam,~H.~J.; Jung,~D.-Y.; Yoo,~W.~J.
  Passivated ambipolar black phosphorus transistors. \emph{Nanoscale}
  \textbf{2016}, \emph{8}, 12773--12779\relax
\mciteBstWouldAddEndPuncttrue
\mciteSetBstMidEndSepPunct{\mcitedefaultmidpunct}
{\mcitedefaultendpunct}{\mcitedefaultseppunct}\relax
\EndOfBibitem
\bibitem[Ryder \latin{et~al.}(2016)Ryder, Wood, Wells, Yang, Jariwala, Marks,
  Schatz, and Hersam]{RWWY2016}
Ryder,~C.~R.; Wood,~J.~D.; Wells,~S.~A.; Yang,~Y.; Jariwala,~D.; Marks,~T.~J.;
  Schatz,~G.~C.; Hersam,~M.~C. Covalent functionalization and passivation of
  exfoliated black phosphorus via aryl diazonium chemistry. \emph{Nat. Chem.}
  \textbf{2016}, \emph{8}, 597--602\relax
\mciteBstWouldAddEndPuncttrue
\mciteSetBstMidEndSepPunct{\mcitedefaultmidpunct}
{\mcitedefaultendpunct}{\mcitedefaultseppunct}\relax
\EndOfBibitem
\bibitem[Ryder \latin{et~al.}(2016)Ryder, Wood, Wells, and Hersam]{RWWH2016}
Ryder,~C.~R.; Wood,~J.~D.; Wells,~S.~A.; Hersam,~M.~C. Chemically tailoring
  semiconducting two-dimensional transition metal dichalcogenides and black
  phosphorus. \emph{ACS Nano} \textbf{2016}, \emph{10}, 3900--3917\relax
\mciteBstWouldAddEndPuncttrue
\mciteSetBstMidEndSepPunct{\mcitedefaultmidpunct}
{\mcitedefaultendpunct}{\mcitedefaultseppunct}\relax
\EndOfBibitem
\bibitem[Qiu \latin{et~al.}(2017)Qiu, da~Jornada, and Louie]{QJL2017}
Qiu,~D.~Y.; da~Jornada,~F.~H.; Louie,~S.~G. Environmental screening effects in
  2D materials: Renormalization of the bandgap, electronic structure, and
  optical spectra of few-layer black phosphorus. \emph{Nano Lett.}
  \textbf{2017}, \emph{17}, 4706--4712\relax
\mciteBstWouldAddEndPuncttrue
\mciteSetBstMidEndSepPunct{\mcitedefaultmidpunct}
{\mcitedefaultendpunct}{\mcitedefaultseppunct}\relax
\EndOfBibitem
\bibitem[Hedin(1965)]{H1965}
Hedin,~L. New method for calculating the one-particle Green's function with
  application to the electron-gas problem. \emph{Phys. Rev.} \textbf{1965},
  \emph{139}, A796\relax
\mciteBstWouldAddEndPuncttrue
\mciteSetBstMidEndSepPunct{\mcitedefaultmidpunct}
{\mcitedefaultendpunct}{\mcitedefaultseppunct}\relax
\EndOfBibitem
\bibitem[Hybertsen and Louie(1986)Hybertsen, and Louie]{HL1986}
Hybertsen,~M.~S.; Louie,~S.~G. Electron correlation in semiconductors and
  insulators: Band gaps and quasiparticle energies. \emph{Phys. Rev. B}
  \textbf{1986}, \emph{34}, 5390\relax
\mciteBstWouldAddEndPuncttrue
\mciteSetBstMidEndSepPunct{\mcitedefaultmidpunct}
{\mcitedefaultendpunct}{\mcitedefaultseppunct}\relax
\EndOfBibitem
\bibitem[Rohlfing and Louie(2000)Rohlfing, and Louie]{RL2000}
Rohlfing,~M.; Louie,~S.~G. Electron-hole excitations and optical spectra from
  first principles. \emph{Phys. Rev. B} \textbf{2000}, \emph{62}, 4927\relax
\mciteBstWouldAddEndPuncttrue
\mciteSetBstMidEndSepPunct{\mcitedefaultmidpunct}
{\mcitedefaultendpunct}{\mcitedefaultseppunct}\relax
\EndOfBibitem
\bibitem[Kharche \latin{et~al.}(2014)Kharche, Muckerman, and
  Hybertsen]{KMH2014}
Kharche,~N.; Muckerman,~J.~T.; Hybertsen,~M.~S. First-principles approach to
  calculating energy level alignment at aqueous semiconductor interfaces.
  \emph{Phys. Rev. Lett.} \textbf{2014}, \emph{113}, 176802\relax
\mciteBstWouldAddEndPuncttrue
\mciteSetBstMidEndSepPunct{\mcitedefaultmidpunct}
{\mcitedefaultendpunct}{\mcitedefaultseppunct}\relax
\EndOfBibitem
\bibitem[Chen and Quek(2018)Chen, and Quek]{CQ2018}
Chen,~Y.; Quek,~S.~Y. Tunable bright interlayer excitons in few-layer black
  phosphorus based van der Waals heterostructures. \emph{2D Mater.}
  \textbf{2018}, \emph{5}, 045031\relax
\mciteBstWouldAddEndPuncttrue
\mciteSetBstMidEndSepPunct{\mcitedefaultmidpunct}
{\mcitedefaultendpunct}{\mcitedefaultseppunct}\relax
\EndOfBibitem
\bibitem[Frimpong and Liu(2021)Frimpong, and Liu]{FL2021}
Frimpong,~J.; Liu,~Z.-F. Quasiparticle electronic structure of two-dimensional
  heterotriangulene-based covalent organic frameworks adsorbed on Au (111).
  \emph{J. Phys.: Condens. Matter} \textbf{2021}, \emph{33}, 254004\relax
\mciteBstWouldAddEndPuncttrue
\mciteSetBstMidEndSepPunct{\mcitedefaultmidpunct}
{\mcitedefaultendpunct}{\mcitedefaultseppunct}\relax
\EndOfBibitem
\bibitem[Neaton \latin{et~al.}(2006)Neaton, Hybertsen, and Louie]{NHL2006}
Neaton,~J.~B.; Hybertsen,~M.~S.; Louie,~S.~G. Renormalization of molecular
  electronic levels at metal-molecule interfaces. \emph{Phys. Rev. Lett.}
  \textbf{2006}, \emph{97}, 216405\relax
\mciteBstWouldAddEndPuncttrue
\mciteSetBstMidEndSepPunct{\mcitedefaultmidpunct}
{\mcitedefaultendpunct}{\mcitedefaultseppunct}\relax
\EndOfBibitem
\bibitem[Thygesen and Rubio(2009)Thygesen, and Rubio]{TR2009}
Thygesen,~K.~S.; Rubio,~A. Renormalization of molecular quasiparticle levels at
  metal-molecule interfaces: Trends across binding regimes. \emph{Phys. Rev.
  Lett.} \textbf{2009}, \emph{102}, 046802\relax
\mciteBstWouldAddEndPuncttrue
\mciteSetBstMidEndSepPunct{\mcitedefaultmidpunct}
{\mcitedefaultendpunct}{\mcitedefaultseppunct}\relax
\EndOfBibitem
\bibitem[Perdew \latin{et~al.}(1996)Perdew, Burke, and Ernzerhof]{PBE1996}
Perdew,~J.~P.; Burke,~K.; Ernzerhof,~M. Generalized gradient approximation made
  simple. \emph{Phys. Rev. Lett.} \textbf{1996}, \emph{77}, 3865\relax
\mciteBstWouldAddEndPuncttrue
\mciteSetBstMidEndSepPunct{\mcitedefaultmidpunct}
{\mcitedefaultendpunct}{\mcitedefaultseppunct}\relax
\EndOfBibitem
\bibitem[Giannozzi \latin{et~al.}(2017)Giannozzi, Andreussi, Brumme, Bunau,
  Nardelli, Calandra, Car, Cavazzoni, Ceresoli, Cococcioni, \latin{et~al.}
  others]{GABB2017}
Giannozzi,~P.; Andreussi,~O.; Brumme,~T.; Bunau,~O.; Nardelli,~M.~B.;
  Calandra,~M.; Car,~R.; Cavazzoni,~C.; Ceresoli,~D.; Cococcioni,~M.
  \latin{et~al.}  Advanced capabilities for materials modelling with Quantum
  ESPRESSO. \emph{J. Phys.: Condens. Matter} \textbf{2017}, \emph{29},
  465901\relax
\mciteBstWouldAddEndPuncttrue
\mciteSetBstMidEndSepPunct{\mcitedefaultmidpunct}
{\mcitedefaultendpunct}{\mcitedefaultseppunct}\relax
\EndOfBibitem
\bibitem[Schlipf and Gygi(2015)Schlipf, and Gygi]{SG2015}
Schlipf,~M.; Gygi,~F. Optimization algorithm for the generation of ONCV
  pseudopotentials. \emph{Comput. Phys. Commun.} \textbf{2015}, \emph{196},
  36--44\relax
\mciteBstWouldAddEndPuncttrue
\mciteSetBstMidEndSepPunct{\mcitedefaultmidpunct}
{\mcitedefaultendpunct}{\mcitedefaultseppunct}\relax
\EndOfBibitem
\bibitem[Berland and Hyldgaard(2014)Berland, and Hyldgaard]{BH2014}
Berland,~K.; Hyldgaard,~P. Exchange functional that tests the robustness of the
  plasmon description of the van der Waals density functional. \emph{Phys. Rev.
  B} \textbf{2014}, \emph{89}, 035412\relax
\mciteBstWouldAddEndPuncttrue
\mciteSetBstMidEndSepPunct{\mcitedefaultmidpunct}
{\mcitedefaultendpunct}{\mcitedefaultseppunct}\relax
\EndOfBibitem
\bibitem[Ambrosch-Draxl \latin{et~al.}(2009)Ambrosch-Draxl, Nabok, Puschnig,
  and Meisenbichler]{ANPM2009}
Ambrosch-Draxl,~C.; Nabok,~D.; Puschnig,~P.; Meisenbichler,~C. The role of
  polymorphism in organic thin films: oligoacenes investigated from first
  principles. \emph{New J. Phys.} \textbf{2009}, \emph{11}, 125010\relax
\mciteBstWouldAddEndPuncttrue
\mciteSetBstMidEndSepPunct{\mcitedefaultmidpunct}
{\mcitedefaultendpunct}{\mcitedefaultseppunct}\relax
\EndOfBibitem
\bibitem[Ismail-Beigi(2006)]{SIB2006}
Ismail-Beigi,~S. Truncation of periodic image interactions for confined
  systems. \emph{Phys. Rev. B} \textbf{2006}, \emph{73}, 233103\relax
\mciteBstWouldAddEndPuncttrue
\mciteSetBstMidEndSepPunct{\mcitedefaultmidpunct}
{\mcitedefaultendpunct}{\mcitedefaultseppunct}\relax
\EndOfBibitem
\bibitem[Deslippe \latin{et~al.}(2013)Deslippe, Samsonidze, Jain, Cohen, and
  Louie]{DSJC2013}
Deslippe,~J.; Samsonidze,~G.; Jain,~M.; Cohen,~M.~L.; Louie,~S.~G. Coulomb-hole
  summations and energies for GW calculations with limited number of empty
  orbitals: A modified static remainder approach. \emph{Phys. Rev. B}
  \textbf{2013}, \emph{87}, 165124\relax
\mciteBstWouldAddEndPuncttrue
\mciteSetBstMidEndSepPunct{\mcitedefaultmidpunct}
{\mcitedefaultendpunct}{\mcitedefaultseppunct}\relax
\EndOfBibitem
\bibitem[Deslippe \latin{et~al.}(2012)Deslippe, Samsonidze, Strubbe, Jain,
  Cohen, and Louie]{DSSJ2012}
Deslippe,~J.; Samsonidze,~G.; Strubbe,~D.~A.; Jain,~M.; Cohen,~M.~L.;
  Louie,~S.~G. BerkeleyGW: A massively parallel computer package for the
  calculation of the quasiparticle and optical properties of materials and
  nanostructures. \emph{Comput. Phys. Commun.} \textbf{2012}, \emph{183},
  1269--1289\relax
\mciteBstWouldAddEndPuncttrue
\mciteSetBstMidEndSepPunct{\mcitedefaultmidpunct}
{\mcitedefaultendpunct}{\mcitedefaultseppunct}\relax
\EndOfBibitem
\bibitem[Adeniran and Liu(2021)Adeniran, and Liu]{AL2021}
Adeniran,~O.; Liu,~Z.-F. Quasiparticle electronic structure of phthalocyanine:
  TMD interfaces from first-principles GW. \emph{J. Chem. Phys.} \textbf{2021},
  \emph{155}, 214702\relax
\mciteBstWouldAddEndPuncttrue
\mciteSetBstMidEndSepPunct{\mcitedefaultmidpunct}
{\mcitedefaultendpunct}{\mcitedefaultseppunct}\relax
\EndOfBibitem
\bibitem[Li and Appelbaum(2014)Li, and Appelbaum]{LA2014}
Li,~P.; Appelbaum,~I. Electrons and holes in phosphorene. \emph{Phys. Rev. B}
  \textbf{2014}, \emph{90}, 115439\relax
\mciteBstWouldAddEndPuncttrue
\mciteSetBstMidEndSepPunct{\mcitedefaultmidpunct}
{\mcitedefaultendpunct}{\mcitedefaultseppunct}\relax
\EndOfBibitem
\bibitem[Tran \latin{et~al.}(2015)Tran, Fei, and Yang]{TFY2015}
Tran,~V.; Fei,~R.; Yang,~L. Quasiparticle energies, excitons, and optical
  spectra of few-layer black phosphorus. \emph{2D Mater.} \textbf{2015},
  \emph{2}, 044014\relax
\mciteBstWouldAddEndPuncttrue
\mciteSetBstMidEndSepPunct{\mcitedefaultmidpunct}
{\mcitedefaultendpunct}{\mcitedefaultseppunct}\relax
\EndOfBibitem
\bibitem[Cudazzo \latin{et~al.}(2012)Cudazzo, Gatti, and Rubio]{CGR2012}
Cudazzo,~P.; Gatti,~M.; Rubio,~A. Excitons in molecular crystals from
  first-principles many-body perturbation theory: Picene versus pentacene.
  \emph{Phys. Rev. B} \textbf{2012}, \emph{86}, 195307\relax
\mciteBstWouldAddEndPuncttrue
\mciteSetBstMidEndSepPunct{\mcitedefaultmidpunct}
{\mcitedefaultendpunct}{\mcitedefaultseppunct}\relax
\EndOfBibitem
\bibitem[Sharifzadeh \latin{et~al.}(2015)Sharifzadeh, Wong, Wu, Cotts, Kronik,
  Ginsberg, and Neaton]{SWWC2014}
Sharifzadeh,~S.; Wong,~C.~Y.; Wu,~H.; Cotts,~B.~L.; Kronik,~L.;
  Ginsberg,~N.~S.; Neaton,~J.~B. Relating the Physical Structure and
  Optoelectronic Function of Crystalline TIPS-Pentacene. \emph{Adv. Funct.
  Mater.} \textbf{2015}, \emph{25}, 2038--2046\relax
\mciteBstWouldAddEndPuncttrue
\mciteSetBstMidEndSepPunct{\mcitedefaultmidpunct}
{\mcitedefaultendpunct}{\mcitedefaultseppunct}\relax
\EndOfBibitem
\end{mcitethebibliography}


\end{document}